\documentclass[aps,prx,reprint,groupedaddress]{revtex4-1}
\usepackage{times}
\usepackage{chemarrow} 
\usepackage{amsmath,amssymb,mathrsfs}
\usepackage{amscd}
\usepackage{graphicx,epsfig}
\usepackage{bbm}
\usepackage{extarrows,chemarrow,xypic} %
\usepackage{hyperref}
\usepackage{microtype}
\usepackage{xcolor}
\DisableLigatures[f]{encoding = *, family = *}
\usepackage[none]{hyphenat}

\def\rd{{\rm d}}

\def\vx{{\bf x}}
\def\vy{{\bf y}}

\begin{document}


\title{Ternary Representation
of Stochastic Change and the Origin of Entropy and Its
Fluctuations}

\author{Hong Qian}
\email[]{hqian@u.washington.edu}
\author{Yu-Chen Cheng}
\email[]{yuchench@u.washington.edu}
\author{Lowell F. Thompson}
\email[]{lfthomps@u.washington.edu}

\affiliation{Department of Applied Mathematics, University of Washington, Seattle, WA 98195, U.S.A.
}

\begin{abstract}
A change in a stochastic system
has three representations: Probabilistic, statistical, and
informational: (i) is based on random variable
$u(\omega)\to\tilde{u}(\omega)$; this induces (ii) the probability distributions $F_u(x)\to F_{\tilde{u}}(x)$, $x\in\mathbb{R}^n$; 
and (iii) a change in the probability measure $\mathbb{P}\to\tilde{\mathbb{P}}$ under 
the same observable $u(\omega)$.  In the {\em informational
representation} a change is quantified by the Radon-Nikodym 
derivative $\ln\big(\frac{\rd\tilde{\mathbb{P}}}{\rd\mathbb{P}}(\omega)\big)=-\ln\big(\frac{\rd F_u}{\rd F_{\tilde{u}}}(x)\big)$ when $x=u(\omega)$. Substituting a random variable into its own density function creates a fluctuating entropy whose expectation has been given by Shannon.  Informational representation of a 
deterministic transformation on $\mathbb{R}^n$ reveals entropic 
and energetic terms, and the notions of configurational entropy of Boltzmann and Gibbs, and potential of mean force of Kirkwood.  
Mutual information arises for correlated $u(\omega)$ and $\tilde{u}(\omega)$; and a nonequilibrium thermodynamic entropy balance 
equation is identified. 
\end{abstract}


\maketitle


\section{Introduction}

	A change according to classical physics is simple:
if one measures $x_1$ and $x_2$ which are traits, in real 
numbers, of a ``same'' type, then $\Delta x=x_2-x_1$
is the mathematical representation of the change;  
$\Delta x\in\mathbb{R}^n$.  How to characterize a
change in a complex world?  To represent a complex,
stochastic world \cite{qian-I}, the theory of probability developed by
A. N. Kolmogorov envisions an abstract space
$(\Omega,\mathcal{F},\mathbb{P})$ called a 
{\em probability space}.   Similar to the Hilbert space underlying
quantum mechanics \cite{pamdirac}, one does not see or touch the
objects in the probability space, $\omega\in\Omega$, nor the 
$\mathbb{P}$.  Rather, one
observes the probability space through functions, say $u(\omega)$,
called random variables which map $\Omega\to\mathbb{R}^n$.
The same function maps the probability measure $\mathbb{P}$
to a cumulative probability distribution function (cdf) $F_u(x)$, $x\in\mathbb{R}^n$.

	Now a change occurs; and based on observation(s) the 
$F_{u}(x)$ is changed to $F_{\tilde{u}}(x)$.  In the current 
statistical data science, one simply works with the two functions 
$F_u(x)$ and $F_{\tilde{u}}(x)$.  In fact, the more complete 
description in 
the {\em statistical representation} is a joint probability 
distribution $F_{u\tilde{u}}(x_1,x_2)$ whose marginal
distributions are $F_u(x_1)=F_{u\tilde{u}}(x_1,\infty)$ 
and $F_{\tilde{u}}(x_2)=F_{u\tilde{u}}(\infty,x_2)$. 

	If, however, one explores a little more on the ``origin'' 
of the change, one realizes that there are two possible sources:
A change in the $\mathbb{P}$, or a change in the $u(\omega)$.
If the $\mathbb{P}\to\tilde{\mathbb{P}}$
while the $u(\omega)$ is fixed,  then according to the
measure theory, one can characterize this ``change of measure''
in terms of a Radon-Nikodym (RN) derivative 
$\frac{\rd\tilde{\mathbb{P}}}{\rd\mathbb{P} }(\omega)$
\cite{kolmogorov-book,ye-qian-18,hqt-19}.  In the rest of this paper, we will assume that all the measures under consideration are absolutely continuous with respect to each other and that all measures on $\mathbb{R}^n$ are absolutely continuous with respect to the Lebesgue measure.  This ensures that all RN derivatives are well-defined.
Note, this is a mathematical object that is defined on the
invisible probability space.  It actually is itself a random variable,
with expectation, variance, and statistics.  

What is the relationship between this {\em informational 
representation} of the change and the observed $F_u$ and 
$F_{\tilde{u}}$?  For $u(\omega),\tilde{u}(\omega)\in\mathbb{R}$,
the answer is \cite{foot.note.RN}:
\begin{equation}
       \frac{\rd\tilde{\mathbb{P}}}{\rd\mathbb{P} }(\omega)
      =   
\left[ \frac{\rd F_u}{\rd F_{\tilde{u}}}\Big(u(\omega) \Big)
           \right]^{-1}.
\label{eq001}
\end{equation}
On the rhs of (\ref{eq001}), $\frac{\rd F_u}{\rd F_{\tilde{u}}}(x)$
is like a probability density function, which is only defined on $\mathbb{R}$.  However, substituting the random variable 
$u(\omega)$ into the probability density function, one 
obtains the lhs of (\ref{eq001}).  Putting a random variable 
back into the logarithm of its own density function to create a 
new random variable is the fundamental idea of fluctuating entropy
in stochastic thermodynamics \cite{qian-pre01,seifert-prl}, and
the notion of self-information \cite{tribus-m,kolmogorov,qian-zgkx}.  
Its expected value then
becomes the Shannon information entropy or intimately related 
relative entropy \cite{hobson}.  The result in (\ref{eq001}) can 
be generalized to $u,\tilde{u}\in\mathbb{R}^n$.  In this case, 
\begin{equation}
\frac{\rd F_u}{\rd F_{\tilde{u}}}\big(x_1,\cdots,x_n\big)
  = \frac{\frac{\partial^n F_u}{\partial x_1 \cdots\partial x_n}  }{\  \frac{\partial^n F_{\tilde{u}}}{\partial x_1 \cdots\partial x_n} \ }.
\end{equation}
In the rest of the paper, we shall consider the $u(\omega)\in\mathbb{R}$.
But the results are generally valid for multidimensional $u(\omega)$.
	
	In this paper, we present key results based on
this informational representation of stochastic change.
We show all types of entropy are unified under the single
theory.  The discussions are restricted on very simple cases; 
we only touch upon the stochastic change with a pair of 
correlated $u(\omega)\to \tilde{u}(\omega)$, which have
respective generated $\sigma$-algebras that are
non-identical in general.  The notion of ``thermodynamic work''
will appear then \cite{hqt-19}.

	The informational and probabilistic representations 
of stochastic changes echo the Schr\"{o}dinger and 
Heisenberg pictures in quantum dynamics \cite{louisell-book}:
in terms of wave functions in the abstract, invisible Hilbert space 
and in terms of self-adjoint operators as observables.

\section{Informational Representation of Stochastic
Change}

{\bf\em Statistics and information: Push-forward and pull-back.}
Consider a sequence of real-valued random 
variables of a same physical origin and their individual cdfs:
\begin{equation}
    F_{u_1}(x), F_{u_2}(x), \cdots, F_{u_T}(x).
\label{0021}
\end{equation}
According to the axiomatic theory of probability built on
$(\Omega,\mathcal{F},\mathbb{P})$, there is a 
sequence of random variables
\begin{equation}
      u_1(\omega), u_2(\omega), \cdots, u_T(\omega),
\label{0020}
\end{equation}
in which each $u_k(\omega)$ maps the 
$\mathbb{P}(\omega)$ to the {\em push-forward measure}
$F_k(x)$, $x\in\mathbb{R}$.  Eq. \ref{eq004} illustrate
this with $u$ and $\tilde{u}$ stand for any $u_i$ and $u_j$.

\vskip -1.2cm
\begin{equation}
\label{eq004}
\begin{picture}(10,60)(10,40)
\put(4,60){$ \mathbb{P}(\omega) $}
\put(75,60){$ F_u(x) $}
\put(4,0){$ \tilde{\mathbb{P}}(\omega) $}
\put(75,0){$ F_{\tilde{u}}(x)  $}
\put(30,63){\vector(1,0){40}}
\put(30,3){\vector(1,0){40}}
\put(20,53){\vector(3,-2){60}}
\put(15,53){\vector(0,-1){40}}
\put(39,66){$u(\omega)$}
\put(39,6){$u(\omega)$}
\put(45,37){$\tilde{u}(\omega)$}
\put(-93,30){$\frac{\rd\tilde{\mathbb{P}}}{\rd\mathbb{P}}(\omega)=\big[ \frac{\rd F_u}{\rd F_{\tilde{u}} }\big(u(\omega)\big) \big]^{-1}$}
\end{picture}
\end{equation}
\vskip 1.7cm\noindent

Informational representation, thus, considers the 
(\ref{0021}) as the push-forward of a sequence of measures $\mathbb{P}_1=\mathbb{P},\mathbb{P}_2,\cdots,\mathbb{P}_T$,
under a single observable, say $u_1(\omega)$.  
This sequence of $\mathbb{P}_k$
can be represented through the fluctuating entropy inside
the $[\cdots]$ below:
\begin{equation}
     \rd\mathbb{P}_k(\omega) =
       \left[ \frac{\rd F_{u_1}}{\rd F_{u_k}}\big(u_1(\omega)\big) \right]^{-1}\rd\mathbb{P}(\omega).
\label{0024}
\end{equation}

Narratives based on information representation have rich 
varieties.  We have discussed above the information representation 
cast with a single, common random variable $u(\omega)$:
$\mathbb{P}_k(\omega)\overset{u}{\longrightarrow}F_k(x)$.
Alternatively, one can cast the information representation
with a single, given $F^*(x)$ on $\mathbb{R}$: $\mathbb{P}_k(\omega)\overset{u_k}{\longrightarrow}F^*(x)$
for any sequence of $u_k(\omega)$.  Actually there is a 
corresponding sequence of measures $\mathbb{P}_k$, whose 
push-forward are all $F^*(x)$, on the real line independent of 
$k$.  Then parallel to Eq. \ref{eq004} we have a
schematic:
\vskip -1.2cm
\begin{equation}
\label{eq07}
\begin{picture}(-10,60)(10,40)
\put(4,60){$ \mathbb{P}(\omega) $}
\put(75,60){$ F_{u_k}(x) $}
\put(4,0){$ \mathbb{P}_k(\omega) $}
\put(75,0){$ F^*(x)  $}
\put(30,63){\vector(1,0){40}}
\put(30,3){\vector(1,0){40}}
\put(20,53){\vector(3,-2){60}}
\put(15,53){\vector(0,-1){40}}
\put(39,66){$u_k(\omega)$}
\put(39,6){$u_k(\omega)$}
\put(45,37){$u_1(\omega)$}
\put(-105,30){$\frac{\rd\mathbb{P}_k}{\rd\mathbb{P}}(\omega)=\Big[ \frac{\rd F_{u_k}}{\rd F^* }\big(u_k(\omega)\big) \Big]^{-1}$}
\end{picture}
\end{equation}
\vskip 1.5cm\noindent

	If the invariant $F^*(x)$ is the uniform distribution, {\em e.g.},
when one chooses the Lebesgue measure on $\mathbb{R}$ with $F^*(x)=x$,  then one has
\begin{equation}
 -\ \mathbb{E}^{\mathbb{P}}\left[\ln\left(\frac{\rd F_{u_k}}{\rd x}\big(u_k(\omega)\big) \right) \right]
\nonumber\\
 	= -\int_{\mathbb{R}} f_{u_k}(x)\ln f_{u_k}(x)\rd x.         
\end{equation}
This is precisely the {\em Shannon entropy}!  More generally
with a fixed $F^*(x)$, one has
\begin{eqnarray}
  && \mathbb{E}^{\mathbb{P}}\left[\ln\left(\frac{\rd F_{u_k}}{\rd F^*}\big(u_k(\omega)\big) \right) \right]
 	=  \int_{\mathbb{R}} f_{u_k}(x)\ln 
  \left[\frac{ f_{u_k}(x)}{f^*(x)}\right] \rd x 
\nonumber\\
	&=& -\int_{\mathbb{R}} f_{u_k}(x)\ln 
    f^*(x) \rd x  +
    \int_{\mathbb{R}} f_{u_k}(x) \ln f_{u_k}(x) \rd x.
\label{0026}
\end{eqnarray}
This is exactly the {\em relative entropy} w.r.t. the stationary
$f^*(x)$.  In statistical thermodynamics, this is called {\em free energy}.  $-\beta^{-1}\ln f^*(x)$ on the rhs of (\ref{0026})
is called {\em internal energy}, where $\beta$ stands for the
physical unit of energy.  The integral is the mean internal energy.

{\bf\em Essential facts on information.}  Several key mathematical
facts concerning the information, as a random variable defined
in (\ref{eq001}), are worth stating.  

First, even though the 
$u(\omega)$ appears in the rhs of the equation, the 
resulting lhs is independent of the $u(\omega)$: It is a 
random variable created from $\mathbb{P}$, $\tilde{\mathbb{P}}$,
and random variable $\tilde{u}(\omega)$, as clearly shown 
in Eq. \ref{eq004}.
This should be compared with a well-known result in elementary
probability: For a real-valued random variable $\eta(\omega)$ and its 
cdf $F_{\eta}(x)$, the constructed random variable $F_{\eta}\big(\eta(\omega)\big)$ is a uniform
distribution on $[0,1]$ independent of the nature of $\eta(\omega)$.

	Second, if we denote the logarithm of (\ref{eq001}) as 
$\xi(\omega)$, $\xi(\omega)=\ln\frac{\rd\tilde{\mathbb{P}}}{\rd\mathbb{P}}$, then one has a result based on the change of
measure for integration:
\begin{subequations}
\begin{eqnarray}
		\mathbb{E}^{\mathbb{P}}\big[\eta(\omega) \big] &=& 
       \int_{\Omega}  \eta(\omega) \rd\mathbb{P}(\omega) \ = \
          \int_{\Omega}  \eta(\omega) 
            \left(\frac{\rd\mathbb{P}}{\rd\tilde{\mathbb{P}}}(\omega)\right)
          \rd\tilde{\mathbb{P}}(\omega)
\nonumber\\
	&=& \int_{\Omega}\eta(\omega)e^{-\xi(\omega)}\rd
    \tilde{\mathbb{P}}(\omega)  
\nonumber\\
      &=& \mathbb{E}^{\tilde{\mathbb{P}}}\left[\eta(\omega) e^{-\xi(\omega)}\right].
\end{eqnarray}
And conversely,
\begin{equation}
	  \mathbb{E}^{\tilde{\mathbb{P}}}\big[\eta(\omega) \big] = 
   \mathbb{E}^{\mathbb{P}}\left[\eta(\omega) e^{\xi(\omega)}\right].
\end{equation}
\end{subequations}
In particular, when the $\eta=1$, the log-mean-exponential 
of fluctuating $\xi$ is zero.  The incarnations of this equality have 
been discovered numerous times in thermodynamics, such as 
Zwanzig's free energy perturbation method \cite{zwanzig}, 
the Jarzynski-Crooks equality \cite{jarzynski,crooks}, and 
the Hatano-Sasa equality \cite{hatano-sasa}.

	Third, one has an inequality,
\begin{equation}
   \ln E^{\mathbb{P}}\big[\xi(\omega)\big] 
        \le \ln E^{\mathbb{P}}\left[e^{\xi(\omega)}\right]  = 0.
\end{equation}
As we have discussed in \cite{hqt-19}, this inequality is 
the mathematical origin of almost all inequalities in connection
to entropy in thermodynamics and information theory. 

	Fourth, let us again consider a real-valued random variable $\eta(\omega)$, with probability density function $f_{\eta}(x)$,
$x\in\mathbb{R}$, and its information, {\it e.g.}, fluctuating entropy $\xi(\omega)=-\ln f_{\eta}\big(\eta(\omega)\big)$ \cite{qian-pre01}. 
Then one has a new measure
$\tilde{\mathbb{P}}$ whose  
$\frac{\rd\tilde{\mathbb{P}}}{\rd\mathbb{P}}(\omega)=e^{\xi(\omega)}$,
and
\begin{eqnarray}
    \int_{x_1<\eta(\omega)\le x_2 }   \rd\tilde{\mathbb{P}}(\omega) &=&
    \int_{x_1<\eta(\omega)\le x_2 }  e^{\xi(\omega)}\rd\mathbb{P}(\omega)
\nonumber\\
   &=& \int_{x_1<\eta(\omega)\le x_2 }  
      \big[ f_{\eta}\big(\eta(\omega)\big) \big]^{-1}
      \rd\mathbb{P}(\omega)
\nonumber\\
	&=&   \int_{x_1}^{x_2} 
            \big( f_{\eta}(y) \big)^{-1} f_{\eta}(y)\rd y
\nonumber\\
   	&=&    x_2-x_1.
\end{eqnarray}
This means that under the new measure $\tilde{\mathbb{P}}$,
the random variable $\eta(\omega)$ has an unbiased 
uniform distribution on the $\mathbb{R}$.
Note that the measure $\tilde{\mathbb{P}}(\omega)$ is 
non-normalizable if $\eta(\omega)$ is not a bounded 
function.  

Entropy is the greatest ``equalizer'' of random variables,
as physical observables inevitably biased!

The forgoing discussion leaves no doubt that 
entropy (or negative free energy) $\xi(\omega)$ is a quantity to be 
used in the form of $e^{\xi(\omega)}$.   This clearly points to 
the origin of partition function computation in statistical 
mechanics.   In fact, it is fitting 
to call the tangent space in the affine structure of the space 
of measures its ``space of entropies'' \cite{hqt-19}; 
which represents the {\em change} of information.

\section{Configurational Entropy in Classical Dynamics}
\label{sec:3}

	By classical dynamics, we mean the representation of dynamical 
change in terms of a deterministic mathematical description,
with either discrete space-time or continuous space-time.
The notion of ``configurational entropy'' arose in this 
context in the theories of statistical mechanics, developed by
L. Boltzmann and J. W. Gibbs, either as Boltzmann's 
{\em Wahrscheinlichkeit} $W$, the Jacobian matrix in a deterministic
transformations \cite{gibbs-book} as a non-normalizable density 
function, or its cumulative distribution.  Boltzmann's entropy emerges 
in connection to macroscopic observables, which are chosen 
naturally from the conserved quantities in a microscopic dynamics.  

	In our present approach, the classical dynamics is a 
deterministic map in the space of observables, $\mathbb{R}^n$.  
The informational representation demands a description of the 
change via a change of measures, and we now show the notion of 
configurational entropy arises.

{\bf\em Information in deterministic change.}
Let us now consider a one-to-one deterministic transformation
$\mathbb{R}\to\mathbb{R}$, which maps the random
variable $u(\omega)$ to $v(\omega)=g^{-1}\big(u(\omega)\big)$,
$g'(x)>0$.  Then
\begin{equation}
   \frac{\rd F_v(x)}{\rd F_u(x)} = 
 \left(\frac{f_u(g(x))}{f_u(x)} \right) g'(x).
\label{equation7}
\end{equation}
Applying the result in (\ref{eq001}) and (\ref{eq004}), 
the corresponding RN derivative
\begin{subequations}
\label{equation8}
\begin{eqnarray}
    -\ln\left[\frac{\rd \tilde{\mathbb{P}}}{\rd \mathbb{P}}(\omega)
    \right]
      &=& \Big[ \ln f_u(g(x)) - \ln f_u(x) +
  \ln g'(x) \Big]_{x=v(\omega)}
\\
     &=& \underbrace{ -\ln f_u\big(v(\omega)\big) +
  \ln \left(\frac{\rd g}{\rd x}\big(v(\omega)\big)\right)}_{\text{ information of $v$ under observable $u$}}  
\\
	&& - 
           \underbrace{ \Big( -\ln f_u\big(u(\omega)\big)  \Big) }_{
           \text{ information of $u$ } }
\end{eqnarray}
\end{subequations}
in which the information of $v=g^{-1}(u)$, under observable
$u(\omega)$, has two distinctly
different contributions: {\em energetic part} 
and {\em entropic part}.   

The energetic part represents the ``changing location'', which characterizes movement in the classical dynamic sense: A point to a point.  It experiences an ``energy'' change where the internal 
energy is defined as $-\beta^{-1}\ln f_u(x)=\varphi(x)$.

The entropic part represents the resolution for measuring
information.  This is distinctly a feature of dynamics in a 
continuous space, it is related to the Jacobian matrix in a 
deterministic transformation: According to the concept of 
Markov partition developed by Kolmogorov and Sinai for chaotic
dynamics \cite{mackey}, there is the possibility of {\em
continuous entropy production} in dynamics with increasing
``state space fineness''.  This term is ultimately related to 
Kolmogorov-Sinai entropy and 
Ruelle's folding entropy \cite{ruelle}. 

{\bf\em Entropy and Jacobian matrix.}
We note that the ``information of $v$ under observable $u$''
\begin{equation}
 -\ln f_u\big(v(\omega)\big) +
  \ln \left(\frac{\rd g}{\rd x}\big(v(\omega)\big)\right)
  \neq  -\ln f_v\big(v(\omega)\big)!
\label{equation9}
\end{equation}
This difference precisely reflects the effect of ``pull-back''
from $\mathbb{R}$ to the $\Omega$ space, there is a breaking 
symmetry between $u(\omega)$ and $v(\omega)$ in
(\ref{equation8}a), when setting $x=v(\omega)$.
Actually, the full ``information entropy change'' associated with
the deterministic map
\begin{equation}
	-\ln f_v\big(v(\omega)\big) + \ln f_u\big(u(\omega)\big) =  
-\ln \left(\frac{\rd g}{\rd x}\big(v(\omega)\big)\right),
\label{eq12}
\end{equation}
as expected.  The rhs is called configurational entropy. 
Note this is an equation between three random variables, 
{\em i.e., fluctuating entropies}, that is valid for all $\omega$.
For a one-to-one map in $\mathbb{R}^n$,
the above $|\rd g(x)/\rd x|$ becomes the absolute value of the
determinant of $n\times n$ Jacobian matrix, which has a 
paramount importance in the theory of functions,
integrations, and deterministic transformations.  
The matrix is associated with an invertible local 
coordinate transform, $y_i = g_i(\vx)$, $1\le i\le n$,
$\vx\in\mathbb{R}^n$:
\begin{equation}
    \frac{\mathscr{D}[y_1,\cdots,y_n]}{\mathscr{D}[x_1,\cdots,x_n]}
 =\left[\begin{array}{ccc}
          \frac{\partial g_1}{\partial x_1} & \cdots & \frac{\partial g_1}{\partial x_n} \\[5pt]
          \vdots & \ddots & \vdots
  \\[5pt]
 \frac{\partial g_n}{\partial x_1}  & \cdots & \frac{\partial g_n}{\partial x_n} 
            \end{array}\right],
\end{equation}
whose determinant is the local ``density change''.  Classical
Hamiltonian dynamics preserves the Lebesgue volume.

{\bf\em Measure-preserving transformation.}
A change with information preservation means the lhs of (\ref{equation8}a) being zero for all $\omega\in\Omega$.  
This implies the function
$g(x)$ necessarily satisfies
\begin{equation}
    -\ln f_u(x) +
  \ln \left(\frac{\rd g(x)}{\rd x}\right)  =
         -\ln f_u\big(g(x)\big).
\label{equation11}
\end{equation}
Eq. \ref{equation11} is actually the condition for 
$g$ preserving the measure $F_u(x)$, with density $f_u(x)$, 
on $\mathbb{R}$ \cite{mackey}:
\begin{subequations}
\begin{eqnarray}
      f_u(x) &=& f_u(g(x))\left(\frac{\rd g(x)}{\rd x}\right)
          =  f_v(x),
\\
	F_u(x) &=& \int_{-\infty}^x f_u(z)\rd z  =  
       F_u\big( g(x)\big).
\end{eqnarray}
\end{subequations}
In this case, the inequality in (\ref{equation9}) becomes an
equality.

	In terms of the internal energy $\varphi(x)$,
then one has
\begin{equation}
	   \beta =
   \frac{1}{\varphi\big(g(x)\big)-\varphi(x)}
 \ln\left(\frac{\rd g(x)}{\rd x}\right),
\end{equation}
which should be heuristically understood as the ratio 
$\frac{\Delta S}{\Delta E}$ 
where $\Delta S=\ln \rd g(x)-\ln \rd x$ and
$\Delta E= \varphi\big(g(x)\big)-\varphi(x)$.   For an infinitesimal
change $g(x)=x+\varepsilon(x)$, we have
\begin{equation}
	   \beta = \frac{\varepsilon'(x)}{\varphi'(x)\varepsilon(x)}.
\end{equation}

{\bf\em Entropy balance equation.}
The expected value of the lhs of (\ref{equation8}a) according 
to measure $\mathbb{P}$ is non-negative.  In fact, consider
the Shannon entropy change associated with
$F_{\tilde{u}}(x) \to F_u(x)$:
\begin{subequations}
\begin{eqnarray}
	&& \int_{\mathbb{R}} f_{\tilde{u}}(x)\ln f_{\tilde{u}}(x) \rd x
     -\int_{\mathbb{R}} f_u(x)\ln f_u(x) \rd x  
\\
	&=& \underbrace{  \int_{\mathbb{R}} f_{\tilde{u}}(x)\ln\left(\frac{f_{\tilde{u}}}{f_u}\right)\rd x }_{\text{ entropy production } \Delta S^{(\text{i})} } 
 + \underbrace{  \int_{\mathbb{R}} 
         \Big[f_{\tilde{u}}(x)- f_u(x)\Big]\ln f_u(x) \rd x
      }_{\text{ entropy exchange } \Delta S^{(\text{e})} }
\nonumber\\
	&=& \mathbb{E}^{\mathbb{P}}
 \left[\ln\left(\frac{\rd \mathbb{P}}{\rd\tilde{\mathbb{P}}}(\omega)\right) \right]
\\
	&&  + \mathbb{E}^{\mathbb{P}}\left[
   \ln\left(\frac{\rd F_u}{\rd x}\right)\big(\tilde{u}(\omega)\big)
   -  \ln\left(\frac{\rd F_u}{\rd x}\right)\big(u(\omega)\big)\right].
\end{eqnarray}
\end{subequations}
This equation in fact has the form of the 
{\em fundamental equation of nonequilibrium thermodynamics} \cite{degroot-mazur,qkkb}:
$\Delta S = \Delta S^{(\text{i})}+\Delta S^{(\text{e})}$. 
The entropy production 
$\Delta S^{\text{(i)}}$ on the rhs is never negative, the entropy
exchange $\Delta S^{\text{(e)}}$ has no definitive sign.  
If $f_u(x)=\frac{\rd F_u(x)}{\rd x}=C$ is a uniform distribution, 
then $\Delta S^{(e)}=0$ and entropy change is the same as entropy
production.

{\bf\em Contracted description, endomorphism, and matrix volume.}
We have discussed above the 
$\mathbb{R}^n\to\mathbb{R}^n$ deterministic invertible 
transformation $\vy=g(\vx)$ and shown that the determinant of 
its Jacobian matrix is indeed the informational entropy change.   
The informational representation, in fact naturally, allows us to 
also consider a non-invertible transformation in the space of 
observable $u(\omega)\in\mathbb{R}^n$ through a
much lower dimensional $g(\vx)\in\mathbb{R}^m$,
$m\ll n$.  In this case, the original $\mathbb{R}^n$ is 
organized by the $m$-dimensional observables, called
``macroscopic'' (thermodynamic) variables in classical 
physics, or ``principle components'' in current data science
and model reduction \cite{kutz-book}.

	The term {\em endomorphism} means that a 
deterministic map  $u\to v=g^{-1}(u)$ in (\ref{equation7}) 
is many-to-one.  In this case, a simple approach is to 
divide the domain of $u$ into invertible parts.  Actually, 
in terms of the probability distributions of $g(\vx)
=(g_1,\cdots,g_m)(\vx)$, the non-normalizable 
density function
\begin{equation}
      W(y_1,\cdots, y_m) =
       \frac{\partial^m}{\partial y_1\cdots\partial y_m}
   \int_{g_1(\vx)\le y_1,\cdots,g_m(\vx)\le y_m} \rd\vx,
\label{eq0019}
\end{equation}
now plays a crucial role in the information representation.  
In fact, the $W$ in (\ref{eq0019}) is the reciprocal of
the matrix volume, {\em e.g.,} the absolute value of the
``determinant'' of the rectangular matrix \cite{ben-israel}
\begin{equation}
\det \left( \frac{\mathscr{D}[y_1,\cdots,y_m]}{\mathscr{D}[x_1,\cdots,x_n]} \right),  
\end{equation}
which can be computed from the product of the singular values of
the non-square matrix.  Boltzmann's {\em Wahrscheinlichkeit}, 
his thermodynamic probability, is when $m=1$.  In that case, 
\begin{equation}
  W(y)=  \frac{\rd}{\rd y}\int_{g(\vx)\le y}\rd\vx
    = \oint_{g(\vx)=y} \frac{\rd\Sigma}{\|\nabla g(\vx)\|},
\end{equation}
in which the integral on rhs is the surface integral on the
level surface of $g(\vx)=y$ .

	One has a further, clear physical interpretation of
$-\ln W(y_1,\cdots,y_m)$ as a {\em potential of
entropic force} \cite{kirkwood-35}, with force: 
\begin{eqnarray}
	\frac{\partial \ln W}{\partial y_{\ell}} &=&
           \frac{1}{W}\frac{\partial W}{\partial y_{\ell}},
 \  \ (1\le\ell\le m).
\label{eq25}
\end{eqnarray}
In the polymer theory of rubber elasticity, a Gaussian
density function emerges due to central limit theorem, 
and Eq. \ref{eq25} yields a three-dimensional Hookean 
linear spring \cite{tlhill}.

\section{Conditional Probability, Mutual Information
and Fluctuation Theorem}

In addition to describing change by $\Delta x = x_2-x_1$,
another more in-depth characterization of a pair of observables
$(x_1,x_2)$ is by their functional dependency $x_2=g(x_1)$,
if any.  In connection to stochastic change, this leads to the
powerful notion of conditional probability, which we now 
discuss in terms of the informational representation.  

First, for two random variables $(u_1,u_2)(\omega)$, the 
conditional probability distribution on $\mathbb{R}\times\mathbb{R}$:
\begin{equation}
   F_{u_1|u_2}(x;y) =  \frac{\mathbb{P}\big\{
   y< u_2(\omega) \le y+\rd y, u_1(\omega)\le x \big\} }{\displaystyle 
     \int_{ y< u_2(\omega) \le y+\rd y} \rd\mathbb{P} }.
\end{equation}
Then it generates an ``informational'' random variable
\begin{equation}
  \xi_{12}(\omega) = \ln\left[ \frac{\ \ \displaystyle \partial F_{u_1|u_2}(x;y)/\partial 
                 x \ \ }{\displaystyle \rd F_{u_1}(x)/\rd x }
      \right]_{x=u_1(\omega),y=u_2(\omega)},
\label{eq23}
\end{equation}
which is a conditional information,
whose expected value is widely know in information
theory as {\em mutual information} \cite{cover-book}:
\begin{eqnarray}
  \mathbb{E}\left[\xi_{12}(\omega)\right]
    &=& \int_{\mathbb{R}^2} f_{u_1u_2}(x,y)
 \ln\left[ \frac{\displaystyle \partial F_{u_1|u_2}(x;y)/\partial 
                 x}{\displaystyle \rd F_{u_1}(x)/\rd x }
      \right]
  \rd x\rd y
\nonumber\\
	&=& \int_{\mathbb{R}^2} f_{u_1u_2}(x,y)
 \ln\left[ \frac{ f_{u_1|u_2}(x;y) }{f_{u_1}(x) }
      \right]
  \rd x\rd y
\nonumber\\
	&=& \int_{\mathbb{R}^2} f_{u_1u_2}(x,y)
 \ln\left[ \frac{ f_{u_1,u_2}(x,y) }{f_{u_1}(x)  f_{u_2}(y)}
      \right]
  \rd x\rd y
\nonumber\\
	&=& \text{MI}(u_1,u_2).
\end{eqnarray}
We note that $\xi_{12}(\omega)$ is actually symmetric
w.r.t. $u_1$ and $u_2$.  The term inside the $[\cdots]$
in (\ref{eq23}
\begin{equation}
    \frac{f_{u_1|u_2}(x;y)}{f_{u_1}(x)} =
      \frac{f_{u_1,u_2}(x,y)}{f_{u_1}(x)f_{u_2}(y)} = 
      \frac{f_{u_2|u_1}(x;y)}{f_{u_2}(x)}.
\end{equation}
In fact the equality $\xi_{12}(\omega)=\xi_{21}(\omega)$ 
is the Bayes' rule.  Furthermore, MI$(u_1,u_2)\ge 0$.  
It has been transformed into a distance function
that satisfies triangle inequality
\cite{dawy05,steuer05}.

In statistics, ``distance'' is not only measured by their
dissimilarity, but also their statistical dependence.
This realization gives rise to the key notion of
{\em independent and identically distributed}
(i.i.d.) random variables.  In the informational
representation, this means $u_1$ and $u_2$ have
same amount of information on the probability
space, {\em and} they have 
zero mutual information.

	Finally, but not the least, for $F_{u_1u_2}(x,y)$, one
can introduce a $\tilde{F}_{u_1u_2}(x,y) = F_{u_1u_2}(y,x)$.
Then entropy production,
\begin{subequations}
\begin{eqnarray}
   \xi(\omega) &=& \ln\left[ \frac{\rd F_{u_1u_2}}{\rd \tilde{F}_{u_1u_2}}(x,y)\right]_{x=u_1(\omega),y=u_2(\omega)}
\\
	&=& \ln\left[ \frac{\frac{\partial^2 F_{u_1u_2}(x,y)}{\partial x\partial y}}{ \frac{\partial^2 F_{u_1u_2}(y,x)}{\partial x\partial y} }\right]_{x=u_1(\omega),y=u_2(\omega)}, 
\end{eqnarray}
\end{subequations}
satisfies the fluctuation theorem \cite{jqq,seifert}:
\begin{equation}
  \frac{\mathbb{P}\{ a<\xi(\omega)\le a+\rd a \}}{\mathbb{P}
      \{-a-\rd a<\xi(\omega)\le -a \} } 
 =  e^a,
\label{eq31}
\end{equation}
for any $a\in\mathbb{R}$.
Eq. \ref{eq31} characterizes the statistical asymmetry 
between $u_1(\omega)$ and $u_2(\omega)$, not independent
in general.  Being identical and independent is symmetric, 
so is a stationary Markov process with reversibility \cite{jqq}.

\section{Conclusion and Discussion}

	The theory of information \cite{cover-book} as a
discipline that exists outside the field of probability and statistics
owes to its singular emphasis on the notion of {\em entropy} 
as a quantitative measure of information.  Our theory shows
that it is indeed a unique {\em representation} of
stochastic change that is distinctly different from, but
complementary to, the traditional mathematical theory of probability.  
In statistical physics, there is a growing awareness of a 
deep relation between the notion of thermodynamic free energy
and information \cite{parrondo}.  The present work clearly
shows that it is possible to narrate the statistical
thermodynamics as a subject of theoretical physics in terms of
the measure-theoretic information, 
as suggested by some scholars who studied deeply 
thermodynamics and the concept of entropy \cite{ben-naim}.  
Just as differential calculus and Hilbert space providing the 
necessary language for classical and quantum mechanics, 
respectively, the Kolmogorovian probability, including the 
informational representation, provides many known results 
in statistical physics and chemistry with a deeper
understanding, such as phase transition and symmetry
breaking \cite{aqtw}, Gibbsian ensemble theory \cite{cwq-19},
nonequilibrium thermodynamics 
\cite{jarzynski,seifert,esposito-review,jqq}, 
and the unification of the theories of chemical kinetics and 
chemical thermodynamics \cite{ge-qian-16}.  In fact, 
symmetry principle \cite{cnyang} and emergent probability 
distribution via limit laws \cite{pwanderson} can both be 
understood as providing legitimate measures {\em a priori},
$\mathbb{P}(\omega)$, for the physical world or the 
biological world.  And stochastic kinematics 
{\em dictates} entropic force and its potential function, the free 
energy \cite{qian-kinematics}. \cite{foot.note.symmetry}

	For a long time the field of information theory and the
study of large deviations (LD) in probability were not 
integrated: A. Ya Khinchin's book was published the same
year as the Sanov theorem \cite{khinchin-book,sanov}.  
Researchers now are agreed upon that Boltzmann's original
approach to the canonical equilibrium energy distribution
\cite{Boltzmann}, which was based on a maximum entropy 
argument, is a part of the contraction of Sanov theorem 
in the LD theory \cite{touchette}.  In probability, the 
contraction principle emphasizes the LD rate function for the 
sample mean of a random variable.\cite{foot.note.contraction}
In statistics and data science, the same 
mathematics has been used to justify the {\em maximum 
entropy principle} which emphasizes a bias, as a 
conditional probability, introduced by observing a sample 
mean \cite{shore-johnson}.  In all these work, Shannon's
entropy and its variant relative entropy, as a single
numerical characteristic of a probability distribution, has 
a natural and logical role. Many approaches 
have been further advanced in applications, {\em e.g.},
surprisal analysis and maximum caliber
principle \cite{levine,dill}.

	The idea that information can be itself a stochastic quantity 
originated in the work of Tribus, Kolmogorov 
\cite{tribus-m,kolmogorov}, and probably many other
mathematically minded researchers \cite{downarowicz}.
In physics, fluctuating entropy and entropy production arose 
in the theory of nonequilibrium stochastic thermodynamics.
This development has significantly deepened the
concept of entropy, both to physics and as the theory of
information.  The present work further illustrates that the
notion of information, together with fluctuating entropy, 
actually originates from a {\em perspective} that is rather 
different from that of strict Kolmogorovian;  with complementarity 
and contradistinctions.  In pure mathematics, the notion of
change of measures goes back at least to 
1940s \cite{cameron-martin}, if not earlier.


\begin{thebibliography}{99}
\bibitem{foot.note.RN}
$\tilde{F}_{u}(x) =$
\begin{eqnarray*}
   &&  \int_{u(\omega)
        \le x}  \rd\tilde{\mathbb{P}}(\omega)
 \  = \   \int_{u(\omega) \le x} 
\frac{\rd F_{\tilde{u}}}{\rd F_u}\big(u(\omega)\big)
 \rd\mathbb{P}(\omega)
\\
	&=&  \int_{-\infty}^x 
\frac{ \left[ \frac{ \rd F_{\tilde{u}}(u) }{\rd u} \right] }{ \left[\frac{ \rd F_u(u) }{\rd u} \right] } f_{u}(u)\rd u
\ = \ \int_{-\infty}^x 
   \left[ \frac{ \rd F_{\tilde{u}}(u) }{\rd u} \right]\rd u
	\  =  \   F_{\tilde{u}}(x).
\end{eqnarray*}
The equality in (\ref{eq001}) is restricted on all the $B\in\sigma\big(u^{-1}(\mathcal{B})\big)\subseteq\mathcal{F}$.
\bibitem{foot.note.symmetry}
There is a remarkable 
logic consistency between fundamental physics pursuing
symmetries in appropriate spaces under appropriate group 
actions and the notion of invariant measures in a theory of 
stochastic dynamics as the legitimate prior 
probability \cite{jaynes,jinwang-rev}.
\end{thebibliography}
\end{document}